\documentclass[preprint,5p,times,authoryear]{elsarticle}
\usepackage{amsmath,pifont,graphicx,amssymb,color,setspace,lineno}
\begin{document}
\begin{frontmatter}
\title{Late veneer and late accretion to the terrestrial planets}
\author[elsi]{R.~Brasser\corref{cor}\fnref{fncrio}}
\cortext[cor]{Corresponding author}
\ead{brasser\_astro@yahoo.com}
\author[cu,has]{S. J. Mojzsis\corref{cor}\fnref{fncrio}}
\ead{mojzsis@colorado.edu}
\author[ceed]{S. C. Werner\fnref{fncrio}}
\author[dun]{S. Matsumura\fnref{fn}}
\author[elsi]{S. Ida}
\address[elsi]{Earth Life Science Institute, Tokyo Institute of Technology, Meguro-ku, Tokyo 152-8550, Japan}
\address[cu]{Department of Geological Sciences, University of Colorado, UCB 399, 2200 Colorado Avenue, Boulder, Colorado 80309-0399, 
USA}
\address[has]{Institute for Geological and Geochemical Research, Research Center for Astronomy and Earth Sciences, Hungarian 
Academy of Sciences, 45 Buda\"{o}rsi Street, H-1112 Budapest, Hungary}
\address[ceed]{The Centre for Earth Evolution and Dynamics, University of Oslo, Sem Saelandsvei 24, 0371 Oslo, Norway}
\fntext[fncrio]{Collaborative for Research in Origins (CRiO)}
\address[dun]{School of Science and Engineering, Division of Physics, Fulton Building, University of Dundee, Dundee 
DD1 4HN, UK}
\fntext[fn]{Dundee Fellow}
\begin{abstract}
It is generally accepted that silicate-metal (`rocky') planet formation relies on coagulation from a mixture of sub-Mars sized 
planetary embryos and (smaller) planetesimals that dynamically emerge from the evolving circum-solar disc in the first few million 
years of our Solar System. Once the planets have, for the most part, assembled after a giant impact phase, they continue to be 
bombarded by a multitude of planetesimals left over from accretion. Here we place limits on the mass and evolution of these 
planetesimals based on constraints from the highly siderophile element (HSE) budget of the Moon. Outcomes from a combination 
of N-body and Monte Carlo simulations of planet formation lead us to four key conclusions about the nature of this early epoch. First, 
matching the terrestrial to lunar HSE ratio requires either that the late veneer on Earth consisted of a single lunar-size impactor 
striking the Earth before 4.45~Ga, or that it originated from the impact that created the Moon. An added complication is that 
analysis of lunar samples indicates the Moon does not preserve convincing evidence for a late veneer like Earth. {Second, the 
expected chondritic veneer component on Mars is 0.06 weight percent.} Third, the flux of terrestrial impactors must have been low ($ 
\lesssim 10^{-6}~M_\oplus$~Myr$^{-1}$) to avoid wholesale melting of Earth's crust after 4.4~Ga, and to simultaneously match the 
number of observed lunar basins. This conclusion leads to an Hadean eon which is more clement than assumed previously. Last, after the 
terrestrial planets had fully formed, the mass in remnant planetesimals was $\sim10^{-3}~M_\oplus$, lower by at least an order of 
magnitude than most previous models suggest. Our dynamically and geochemically self-consistent scenario requires that future N-body 
simulations of rocky planet formation either directly incorporate collisional grinding or rely on pebble accretion.
\end{abstract}
\end{frontmatter}
\section{Introduction}
\label{sec:int}
The formation of the terrestrial planets is a long-standing problem that is gradually being resolved. In traditional ㄑdynamical models 
the terrestrial planets grow from a coagulation of planetesimals into protoplanets and subsequently evolve into a giant impact phase, 
during which the protoplanets collide with each other to lead to the terrestrial planets. Several variations of this scenario exist, of 
which the {\it Grand Tack} model is currently popular \citep{W11}. The Grand Tack relies on early gas-driven migration of Jupiter and 
Saturn to gravitationally sculpt the inner solid circum-solar disc down to $\sim$1~AU after which terrestrial planet formation proceeds 
from solids in an annulus ranging from roughly 0.7~AU to 1~AU. Grand Tack has booked some successes, such as its ability to reproduce 
the mass-orbit distribution of the terrestrial planets, the compositional gradient, and total mass of the asteroid belt \citep{W11}. 
Subsequent evolution of the solar system after terrestrial planet formation, all the way to the present, however, has mostly been 
studied in separate epochs with disconnected simulations. \\

\citet{B16} scrutinised the Grand Tack model in more detail { and built a database of simulations that is used here}. Since 
published simulations had rarely been run for much longer than 200~Myr into the evolution of the solar system, we sought to test the 
model predictions specific to the long-term evolution of the terrestrial system. In this work we calculate the { evolution of the 
terrestrial planets for up to 300~Myr. We aim to obtain the} amount of mass accreted by the terrestrial planets after the Moon-forming 
event and whether this accreted mass is compatible with { the highly-siderophile element (HSE) budgets of the inner planets}, the 
early lunar and terrestrial cratering records and the nature of the purported late veneer. We also consider the surface conditions on 
the Hadean Earth from geochemical data and conclude with the implications of our simulations for future models of terrestrial planet 
formation.

\section{Constraints from the Moon on the remnant planetesimal mass}
The unexpectedly high abundance of HSEs in Earth's upper mantle is a mystery because it is expected that these elements would be 
effectively sequestered into the core. One popular explanation suggests that Earth accreted a further 0.5-0.8 weight percent (wt\%) of 
its mass after core separation and after the Giant Impact (GI) that formed the Moon \citep{W09}. The dearth of lunar mantle HSEs 
indicate that the Moon accreted approximately 0.02-0.035 wt\% \citep{DW15,K15}. The ratio of accreted mass between Earth and the Moon 
is then $1950 \pm 650$, which is curious because the ratio of the gravitational cross sections of the Earth and Moon is far less 
($\sim$20).\\

{ There is substantial debate in the literature about a possible late veneer on Mars. Osmium isotopes in martian meteorites 
indicate that Mars accreted chondritic material after core formation \citep{Brandon12}, although it is unclear how much material was 
added to the martian mantle.} \citet{W09} suggested that Mars experienced a mass augmentation of $\sim$0.7 wt\%, comparable to 
Earth's. 
A recent analysis of metal-silicate partitioning for the platinum group elements in martian meteorites, however, combined with 
theoretical partitioning models used to construct inverse models of Mars' mantle composition, instead show that the concentration of 
HSEs in the martian mantle can be solely established by metal-silicate equilibration early in the planet's history \citep{Ri15}. This 
obviates the need for substantial accretion on Mars during the late veneer epoch. Effective removal of the requirement for 
much accretion on Mars is important because now the Moon need no longer be regarded as anomalous in the relatively low amount of 
material it accreted after its formation; only Earth's unusually high HSE abundance demands explanation. This allows us to predict the 
amount of accretion experienced by Mars when calibrated to the Moon. \\ 

The high ratio of the terrestrial and lunar HSE budgets led \citet{B10} to conclude that the size-frequency distribution of the 
remaining planetesimals from planet formation had to have been shallow even at large sizes, and the majority of the mass delivered to 
the Earth should have come from a few large objects comparable to Ceres. The low number of objects leads to a stochastic impact regime 
for large objects that statistically favours collisions with Earth \citep{S89}. Hence, the amount of mass accreted on the Moon must be 
representative of the mass in remnant planetesimals from terrestrial planet formation { that are volumetrically smaller than those 
colliding with Earth.}\\

The probability that the mass of each impactor exceeds $m$ is $P(m) = (m/m_{\rm min})^{-\gamma}$, with $m>m_{\rm min}$. When $\gamma 
\le 1$ the total delivered mass is dominated by a single projectile, and the approximate mass delivered to the Earth versus the Moon 
is $m_{\rm e}/m_{\rm l} = (\sigma_{\rm e}/\sigma_{\rm l})^{1/\gamma}$ \citep{S89}, where $\sigma_{\rm e,l}$ are the gravitational 
cross sections of the Earth and Moon, respectively. { This ratio can become high when $\gamma \ll 1$. Since $\sigma_{\rm 
e}/\sigma_{\rm l} \sim 20$, the probability of the Earth being struck by the largest 13 objects is $(19/20)^{13} >0.5$. The largest 
verifiable projectile to have struck the Moon created the South Pole-Aitken basin; its diameter was $\sim$170~km 
\citep{Pot12}. The collision probability of planetesimals with the Earth is 12\% (see below), so there are a total of 100 objects with 
$D>170$~km, and thus the expected diameter of the largest projectile (99$^{\rm th}$ percentile) is approximately 1700~km, assuming a 
main asteroid belt size-frequency distribution ($\gamma \sim 0.7$).}\\

Recently \citet{R13} showed that a population of planetesimals with a total mass $\sim$0.05~$M_\oplus$~could reproduce the HSE 
signature in Earth's mantle after the GI over the next few hundred million years. Here we argue that the duration of the time interval 
that allowed for the mixing of HSEs into the mantle must have been considerably shorter, depending on the exact timing of the 
Moon-forming event, and should have mostly finished near 4.42~Ga, roughly $\sim$150~Myr after the formation of the proto-Sun at 
4.57~Ga. Our arguments for this duration are as follows.\\

Radiogenic dating of lunar Apollo samples and meteorites indicate that lunar crust formation was well under way by 4.42~Ga 
\citep{Nem09}, which is the age of the oldest lunar zircon thus far documented. Recent analysis of zircons in martian meteorite NWA 
7533 indicates that the earliest crust on Mars formed before 4.43~Ga \citep{H13}. Taken together these ages suggest that both 
the Moon and Mars had begun to form a crust some time before 4.42~Ga. In this work, we refer to {\it late accretion} as those impacts 
that occurred after continual preservation of the planetary crust. An impact large enough to have destroyed most of the crust and 
contaminate planetary mantles with HSEs is dubbed a {\it late veneer}. We shall designate the time interval between the Moon-forming 
giant impact (GI) near 4.5~Ga and crust formation at 4.42~Ga as the {\it late veneer epoch} and the time interval after 4.42~Ga as the 
{\it late accretion epoch}. Late accretion impacts are inefficient at mixing any HSEs into the mantle and thus the majority of late 
veneer impacts, which delivered the HSEs, must have occurred before crust formation. The end of the late veneer epoch corresponds 
closely to the last major differentiation event on the Earth at 4.45~Ga \citep{All08}, during which the bulk silicate reservoirs were 
separated when the crust was still molten.\\

The above arguments can be used to constrain the mass of leftover planetesimals in the inner solar system subsequent to the 
Moon-forming event, delimiting the rate at which this material is cleared by the terrestrial planets. In principle the total mass 
of planetesimals decreases with time as $m_{\rm left} = m_{\rm init}f(t)$, where $f(t)$ is the decay function. The amount of 
mass accreted on the Moon between the time of the GI, assumed to have occurred at 4.5~Ga, and crust formation is 

\begin{equation}
m_{\rm init}P_{\rm l}[f(140)-f(60)]=0.025\, {\rm wt}\% \sim 3 \times 10^{-6}~M_\oplus,
\label{eq:maccr}
\end{equation} 
{ where is $m_{\rm init}$ is the initial mass in planetesimals and $P_{\rm l}$ is the probability of impact of a planetesimal with 
the Moon. The accretion is measured from 4.5~Ga to 4.42~Ga, the approximate duration of the late veneer epoch in our model}. Equation 
(\ref{eq:maccr}) holds regardless of the size-frequency distribution of the remnant planetesimals.\\

In a collisionless system, the mass in leftover planetesimals follows a stretched exponential decay $m_{\rm left} = m_{\rm 
init}\exp[-(t/\tau)^\beta]$. \citet{B16} find { a stretching parameter} $\beta \sim 0.44$, and { e-folding time} $\tau \sim 12 
$~Myr. { Assuming that this rate of decay holds until the birth of the solar system, substituting 0.5\% for the impact probability 
with the Moon (see below)} the mass of remnant planetesimals at the time of the GI is then $m_{\rm left} = 10^{-3}$~$M_\oplus$. \\

This mass is at least an order of magnitude lower than the 0.05~$M_\oplus$ that \citet{B16} obtained with the canonical Grand Tack 
model, or most other terrestrial planet formation simulations. { For example, the classical simulations of \citet{Mat16} also have a 
remnant planetesimal mass of 0.05~$M_\oplus$ after 200~Myr.} If the estimate of the remaining mass in planetesimals from 
\citet{B16} is correct, then after 150~Myr of evolution it is typically 5\% of the initial mass. At the time of the Moon-forming 
impact 
at 4.5~Ga the planetesimal mass typically is $m_{\rm left}\sim 0.1$~$M_\oplus$. Then the { Moon is expected to be struck by 
$0.1\times 0.005\times[f(140)-f(60)]=4.2 \times 10^{-5}$~$M_\oplus$ $\sim 0.35$ wt\%, in contradiction with its HSE budget}. In 
conclusion, the mass in remnant planetesimals {\it must} have been low ($\lesssim 10^{-3}$~M$_\oplus$) at the time the Moon formed.\\

The above analysis assumes that the mass accreted by the Moon is mostly in the form of small planetesimals so that the accretion can 
be considered { perfect and nearly continuous}. These assumptions may not be entirely correct \citep{B10,R13} and we need to better 
determine the total accreted mass on each planet. { To verify the above claims we ran a series of N-body simulations of the solar 
system continuing from \citet{B16}, followed by} Monte Carlo impact experiments that rely on the impact probabilities { from the 
N-body simulations} to determine the total mass of leftover planetesimals, and the relative amount of mass accreted by the Earth, the 
Moon and Mars for a given size-frequency distribution.

\section{N-body simulations of post-formation evolution: Setup and method}
\label{sec:nm}
We placed the terrestrial planets at their current semi-major axes with their current masses but with much lower initial 
eccentricities and inclinations { that are typical outcomes in \citet{B16}}. These will increase later during giant planet 
migration \citep{B13}. We placed the gas giants on nearly-circular nearly-coplanar orbits using the configuration of \citet{T05}.\\

\begin{figure}[t]
\resizebox{\hsize}{!}{\includegraphics[angle=-90]{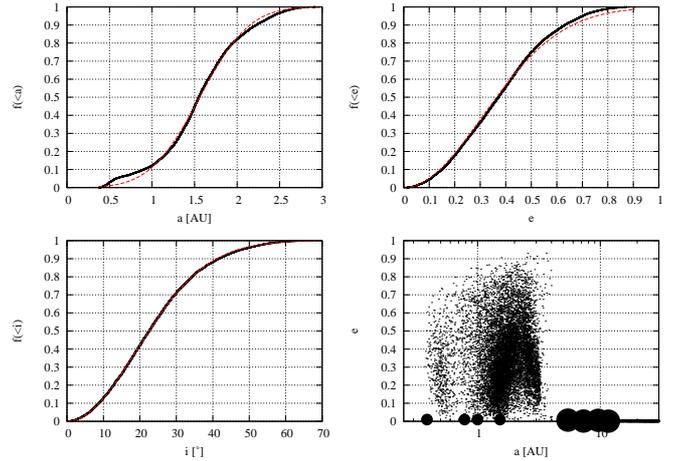}}
\caption{Cumulative distributions (black) and their best fits (red) of the semi-major axis (top, left), eccentricity (top, right), 
inclination (bottom, left) of the planetesimals that were left after terrestrial planet formation simulations in the framework of the 
Grand Tack by \citet{B16}. The bottom-right panel shows the distribution of planetesimals in semi-major axis ($x$) and 
eccentricity ($y$), and the positions of the planets.}
\label{fig:lhbinit}
\end{figure}

The initial orbits of the planetesimals were generated from their orbital distribution from the Grand Tack simulations. Even though 
these include planetesimals from simulations that did not successfully reproduce the terrestrial planets, we concluded that removing 
these had little influence on the final distribution. Only planetesimals that crossed at least one terrestrial planet but not 
Jupiter were considered { since the latter would quickly eject them.} Planetesimals were generated by randomly picking a semi-major 
axis, eccentricity and inclination from their respective distributions { that were fitted with simple functions}, and { imposing 
the condition that the maximum perihelion and aphelion distances satisfy} $q<1.5$~AU and $Q<4.5$~AU (see Supplement). No planetesimal 
was allowed to be entirely inside of the orbit of Mercury. The other three orbital elements are all angles that were chosen uniformly 
random from 0 to 360$^\circ$. With these initial orbits 62\% of planetesimals are Earth crossers, 34\% cross Venus and just 8\% are 
Mercury crossers. By selection all { have their perihelion inside} Mars' orbit, { and the semi-major axis distribution peaks 
just beyond Mars' orbit. The fits and initial conditions are shown in Fig.~\ref{fig:lhbinit}}. \\

{ We impose a size-frequency distribution for the planetesimals so that we have a more realistic amount of mass accreted onto a 
planet than with equal-mass planetesimals.} Following \citet{R13} our model used a cumulative power-law size distribution with a slope 
of either 1 or 2 and a planetesimal radius between 250~km and 2000~km or 250~km to 1000~km. A total mass in planetesimals uniformly 
random from 0.023~$M_\oplus$ to 0.077~$M_\oplus$, bracketing the typical end values of \citet{B16}, was assumed. The mass of each 
planetesimal was computed using a density of 3000~kg~m$^{-3}$. We ran 16 simulations for each combination of power law slope and 
maximum size.\\

All simulations included the gas giants, terrestrial planets and planetesimals, and were integrated with the symplectic SyMBA 
package \citep{Dll98} for 150~Myr, covering the epoch from 4.4~Ga to 4.25~Ga, with a time step of 3.65 days. { The planetesimals 
only gravitationally interact with the planets but not with each other.} Planetesimals were removed once they were farther than 50~AU 
from the Sun or when they collided with a planet or ventured closer than 0.1 AU to the Sun. { We assume perfect accretion, but 
ultimately this does not appear to make a large difference \citep{R13}.}

\section{Results from N-body simulations}
The fraction of remaining planetesimals decreases as a stretched exponential $f = \exp[-(t/\tau)^\beta]$ with $\beta=0.85\pm 0.07$ and 
$\beta \log \tau = 1.6 \pm 0.15$. This decay law is similar to a roughly exponential function with e-folding time 100~Myr. After 
300~Myr of Solar System evolution (150~Myr from the simulations of \citet{B16} and a further 150~Myr here) there is still about 1.5\% 
of the total mass left in remnant planetesimals. This result does not account for their collisional evolution. \\

It appears that Mars is the bottleneck in eliminating this population on long time scales. Encounters with Mars are weak so that { 
the transfer time to} Earth-crossing orbits is long. This slow removal of Mars crossers is a property of the Solar System, and is the 
most likely reason for the initial semi-major axis distribution of the planetesimals peaking near the orbit of Mars.\\

\begin{table}
\begin{tabular}{ccccc}
Quantity & Mercury & Venus & Earth & Mars \\ \hline \\
Mass (wt\%) & 2.7 $\pm$ 4.5 & 0.8 $\pm$ 0.5 & 0.7 $\pm$ 0.5 & 1.4 $\pm$ 2.7 \\
$P_{\rm imp}$ [\%]& 4.0 $\pm$ 0.9 & 18.0 $\pm$ 1.6 & 12.8 $\pm$ 1.8 & 1.9 $\pm$ 0.7 \\
$v_{\rm imp}$ [km s$^{-1}$] & 34.2 $\pm$ 13.6 & 24.3 $\pm$ 7.5& 21.1 $\pm$ 6.2& 13.4 $\pm$ 5.1
\end{tabular}
\caption{The fractional amount of late accreted mass (in wt\%), mean impact probability and mean impact speed of projectiles with 
each planet from the N-body simulations performed here.}
\label{tab:gtsnr}
\end{table}
In Table~\ref{tab:gtsnr} we list the { fractional amount of mass added to each planet}, the mean collision probability { and 
mean impact speed} of a planetesimal with each planet. The impact probability is computed as the number of planetesimals that collide 
with a terrestrial planet divided by the total number of planetesimals at the start of the simulation (see Supplement). The reported 
uncertainties herein are the standard deviations { between different simulations}. The values computed with this method are 
consistent with the previous counts within uncertainties; the average probability of colliding with the Moon is 0.53\% and the average 
lunar impact speed is 17~km~s$^{-1}$. \\

{ Our results indicate that during the late veneer and late accretion epochs the terrestrial planets all accrete a comparable 
amount of their total mass.} To determine whether the remaining planetesimal mass and clearing rate are 
physically plausible, their removal through impacts needs to be compared to the lunar cratering record and constraints imposed by 
geochronology. Better statistics are needed of the nature of the impacts on each planet by taking into account the size-frequency 
distribution of the projectiles and impact velocities. { To address this we turn to a series of Monte Carlo impact experiments, 
described below.}

\section{Monte Carlo impact simulations}
We calculate the diameter of a random planetesimal between 1~km and 2000~km { with a size-frequency distribution that matches the 
main asteroid belt.} It was assumed that each planetesimal has a density of 3000~kg~m$^{-3}$ and its mass is added to the total mass. 
We determine whether the planetesimal will collide with either Earth or Mars using the impact probabilities listed on the second row 
of Table~\ref{tab:gtsnr}, assuming that these are valid during the late veneer epoch; it is noteworthy that the probability of 
colliding with the Moon is barely affected by lunar tidal evolution { because the Moon's gravitational focusing is negligible}. If 
the planetesimal strikes one of these three bodies, its mass is added to the total mass accreted by the target because we assume 
perfect accretion for simplicity. We continue to generate planetesimals until the Moon has accreted 0.025 wt\%. Our runs are 
time-independent, so we used them to check accretion during both the late veneer epoch and subsequent late accretion epoch. During the 
late accretion epoch the accreted mass is halved: applying the same methods and decay law from Section~2 we calculate that the total 
mass impacting the Moon after the late veneer epoch and until 4.25~Ga is 0.013 wt\%.\\

\begin{figure}[t]
\resizebox{\hsize}{!}{\includegraphics[angle=-90]{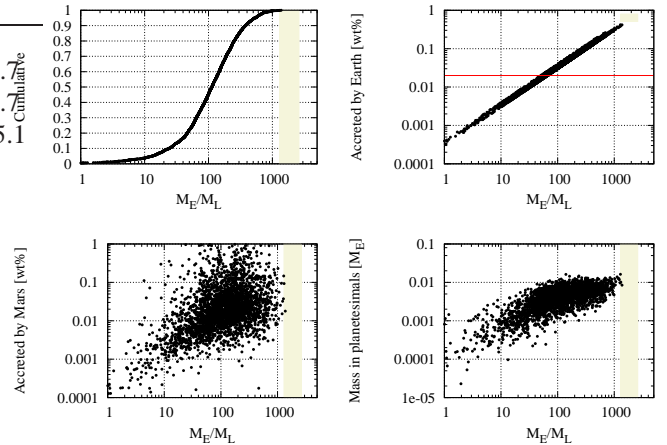}}
\caption{Results of Monte Carlo impact simulations. The $x$-axis is always the ratio of mass accreted by Earth versus that of the 
Moon. The range in the ratio implied from the HSE element abundance is highlighted in beige, and is taken as $1950 \pm 650$. Top-left: 
Cumulative fraction of ratio of accreted mass. Top-right: total mass fraction accreted by the Earth (wt\%). The red line is the 
threshold for wholesale melting of the crust. Bottom-left: total mass fraction accreted by Mars (wt\%). Bottom-right: total mass in 
remnant planetesimals ($M_\oplus$).}
\label{fig:mcimpacts}
\end{figure}

The results for the late veneer epoch are shown in Fig.~\ref{fig:mcimpacts}. The top-left panel displays the cumulative fraction 
of the ratio of mass accreted by the Earth and the Moon. The beige band indicates the range from HSE data. It is evident that the 
probability of matching the observed high ratio of accreted mass is below 1\%, and thus reproducing this high amount of relative 
accretion between the Earth and the Moon { from leftover planetesimals with diameters $D<2000$~km is unlikely}. The top-right panel 
shows the amount of mass accreted by Earth. The bottom-left panel displays the fractional amount of mass accreted by Mars. This is 
generally of the same order as the fractional amount accreted by the Earth: the mean mass accreted by Earth is 0.05$\pm$0.04 wt\% and 
for Mars is 0.06$\pm$ 0.03 wt\%. Last, the bottom-right panel shows the total mass in remnant planetesimals from terrestrial planet 
formation. The mean total mass in planetesimals at the time of the Moon-forming impact is typically 0.0046~$M_\oplus$, lower than what 
was found in \citet{B16}, but still higher than what is predicted by the analysis in Section~2.\\

\begin{figure}[t]
\resizebox{\hsize}{!}{\includegraphics[angle=-90]{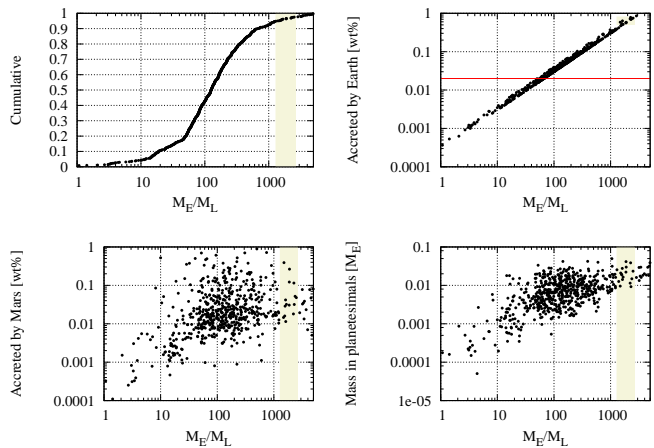}}
\caption{Same as Fig.~\ref{fig:mcimpacts} but in this case the maximum diameter of the planetesimals is increased to 4000~km.} 
\label{fig:mcimpactsq4}
\end{figure}

The results in Fig.~\ref{fig:mcimpacts} are cause for concern. { Our setup rejects the observed HSE ratio between the Earth 
and Moon during the late veneer epoch at $>99\%$ confidence. \citet{Mar14} were unable to reproduce the HSE ratio with a maximum 
plantesimal diameter of 1000~km. As such, we increased the maximum planetesimal diameter to 4000~km in like manner. The results are 
displayed in Fig.~\ref{fig:mcimpactsq4}. It is now possible to reproduce the terrestrial to lunar HSE ratio at the 5\% level and the 
average total mass colliding with Earth is 0.11 $\pm$ 0.22 wt\%, for Mars it is 0.15 $\pm$ 0.34 wt\% and the total leftover 
planetesimal mass is 0.01~$M_\oplus$.\\

We subsequently ran both models for the late accretion epoch, during which the Moon is expected to accrete an additional 0.013 wt\%. 
The results are similar to what is displayed in Fig.~\ref{fig:mcimpacts} but the Earth now only accretes an average of 0.03 $\pm$ 0.03 
wt\% and Mars 0.03 $\pm$ 0.02 wt\% and the mass in leftover planetesimals is approximately 0.003~$M_\oplus$; these values increase to 
0.05 $\pm$ 0.13 wt\%, 0.03 $\pm$ 0.17 wt\% and 0.05~$M_\oplus$ when the largest planetesimal's diameter is 4000~km.}\\

{ Even though increasing the largest planetesimal diameter allows us to reproduce the measured HSE ratio, it is likely that the 
intensity of the bombardment and the high number of large ($D>1000$~km) objects striking the Earth violates geochemical 
and geochronological constraints. \citet{AB13} state that an addition of 0.03 wt\% over 100~Myr will cause $\sim$ 50\% of Earth's 
crust 
to experience melting, while \citet{Mar14} claim that the intensity of their early bombardment causes a cumulative fraction of 
100-400\% of Earth's crust to be buried by impact melt.}\\

A major episode of post-late veneer wholesale crustal melting is inconsistent with the geochronology and there are multiple lines of 
evidence that support this claim. First, the last major differentiation event of the Earth's crust occurred near 4.45~Ga \citep{All08} 
(cf. \citet{Roth13}). This claim is supported by recent measurements of deep-seated chondritic xenon in the mantle that indicate its 
source has been isolated from the rest of the mantle around 4.45 Ga \citep{Ca16}. Second, there are terrestrial zircons dating back to 
4.38~Ga \citep{V14}, which would not have survived such an intense bombardment. Third, Ti-in-zircon thermometry results by 
\citet{Wi12} 
conclude that zircon derived from impactites do not represent a dominant source for the Hadean grains thus far documented from Western 
Australia. Fourth, \citet{C14} show from Ti-in-zircon thermometry analysis that Hadean zircons generally formed in a lower-temperature 
environment than Icelandic mid-ocean ridge basalt; and finally, the Monte Carlo experiments of \citet{Mar14} indicate that the Hadean 
crust should all be gone, inconsistent with the oldest age of terrestrial zircons, and slightly older zircons from the Moon and Mars 
\citep{Nem09,H13}. Taken together, the geochronology indicates that the amount of impacted mass on the Hadean Earth had to be low, and 
direct lines of evidence mitigate a late veneer on Earth after ca. 4.45~Ga \citep{F16}.\\

We are faced with two incompatible constraints: on the one hand we want the accreted mass on the Earth and Mars to be low while at the 
same time we want the Earth to accrete enough to explain the terrestrial to lunar HSE ratio. One way to solve this dilemma is to 
assume that the accretion was not as stochastic as displayed in Fig.~\ref{fig:mcimpacts}. The most parsimonious way to match these 
constraints simultaneously is if the event that delivered the HSEs to the Earth's upper mantle after core formation was from a 
singular late veneer impact unique to Earth, followed by very little accretion afterwards from the background flux of remaining 
planetesimals. The idea of the late veneer from a small number of big objects is not new \citep{B10}, but our analysis leads us to 
conclude it was most likely a single impact, rather than several, and certainly not a multitude. { This would also most likely 
result in a low amount of accretion ($\lesssim 0.1$ wt\%) on Mars after 4.5~Ga.}

\section{Implications and predictions}
\subsection{The Late Veneer as a singular event}
If the late veneer on the Earth is caused by a single impact, a roughly lunar-sized impactor striking Earth would shatter its core and 
cause its HSEs to be suspended in the terrestrial mantle. Conceptually, a singular late veneer event to Earth is not 
inconsistent with the geochemistry, and is supported by the evolution of tungsten isotopes in the terrestrial mantle. \citet{Wb11} 
state that the pre-veneer terrestrial mantle has an excess of $^{182}$W compared to the current mantle of $\varepsilon 
^{182}$W$=0.15-0.2$, while the current mantle sits at 0. \citet{Wb11} conclude that approximately 0.8 wt\% of Earth's mass had to have 
been added in chondritic proportions to offset this excess $^{182}$W; this estimate is compatible with \citet{W09}. The chondritic 
tungsten isotopes from such a `stray dog' impactor would then subsequently be homogenised in the mantle on a time scale of about 1~Gyr 
\citep{Ma09,F16}; we add that a similar process can be traced by $\varepsilon ^{142}$Nd \citep{Roth13}. During the GI the impactor's 
core is thought to merge with Earth's through the terrestrial magma ocean\citep{CA01}. In this models of \citet{Wb11,Wb15} this 
process preserves the pre-late veneer mantle with $\varepsilon^{182}$W$=+0.15$. In the late veneer epoch a single lunar-sized object 
struck the Earth and augmented its mass with chondritic material ($\varepsilon^{182}$W$ \sim -2$) to the pre-late veneer mantle. 
Convective processes in the `upper’ mantle mixed this late veneer material to ultimately homogenise the mantle. By ca. 3300 Ma 
top-to-bottom homogenisation of the late veneer material is mostly complete in the mantle except for lower mantle domains, which are 
expected to have preserved a pre-late veneer $\varepsilon^{182}$W signature. Modern-day plate tectonics and mantle convection will do 
the rest to homogenise the upper mantle in $\varepsilon^{182}$W \citep{Wb15}. { The Late Veneer mechanism is not universally 
accepted, however, and alternative interpretations of HSE in the mantle have been presented.}\\

A late veneer consisting of a steady flux of small impactors will not reproduce the HSE ratio because then the terrestrial 
to lunar mass accretion should be comparable to their gravitational cross sections, which geochemical data show is not the case.\\

{ The origin of this `stray dog' embryo is difficult to trace. Since Jupiter clears about 99.5\% of the planetesimals beyond 1~AU 
in the nominal Grand Tack model, it is likely the embryo originated from closer to the Sun or was pushed inwards by Jupiter and 
subsequently rattled around the inner solar system until it collided with Earth. Alternatively there were originally more than four 
terrestrial planets and that the system of multiple planets became unstable in the first 100~Myr. The unstable fifth planet then 
subsequently collided with the Earth in the form of the late veneer.\\

Such a late impact may also account for the inclination offset between the lunar orbit and the Earth's spin. A grazing impact near 
Earth's pole with a lunar-sized object at escape velocity changes Earth's obliquity by approximately 

\begin{equation}
\delta \varepsilon = (m/Cm_\oplus)(Gm_\oplus/R_\oplus^3)/\nu^2 \sim 15^\circ,
\end{equation}
where $C$ is normalised the moment of inertia and $\nu$ is the rotation rate. In the equation above we used a rotation period of 14~h 
and $C=0.33$. Such an impact obviates the need for many close encounters between the Moon and remnant planetesimals \citep{PM15}.\\

In principle, the Earth's HSE signatures could also be generated from the impactor's core that produced the Moon. Some impactor 
material formed into the lunar core with most being suspended in the Earth's upper mantle \citep{NT89,S16} because of Earth's stronger 
gravity \citep{Kraus15}. The benefit of this approach is that it does not require a second large impact on Earth. The resulting HSE 
ratio can then be arbitrarily large or small, and thus reproducing the HSE ratio becomes a matter of the exact nature of the impact. 
We do note that most Moon-forming impact scenarios yield a molten target Earth that results in merger of the impactor's core with that 
of the proto-Earth\citep{CA01}. Under these conditions, the HSEs should have been effectively stripped from Earth's mantle and 
segregated. In this study, however, we cannot rule in favour of either scenario.}

\subsection{Lunar cratering record}
A dynamical model of terrestrial planet formation should be able to quantitatively reproduce the number of known ancient lunar 
craters, including the basins, { over the time span of a few hundred million years. Here we test our Monte Carlo simulations 
against the lunar basin record and cratering chronology, using the established work of \citet{Neu01} and the recent model of 
\citet{W14}}. \\

\begin{figure}[t]
\resizebox{\hsize}{!}{\includegraphics[angle=-90]{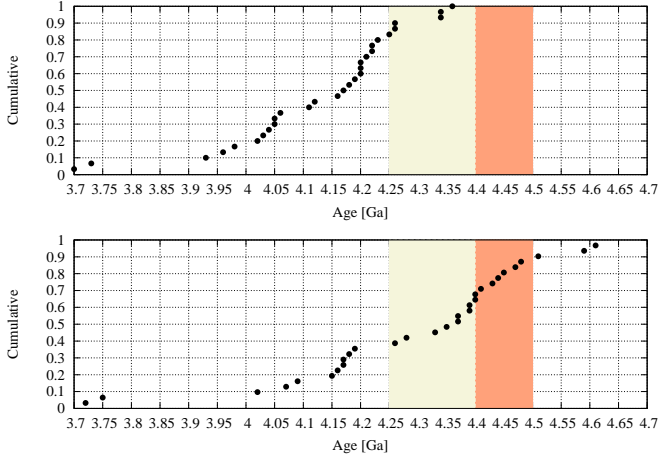}}
\caption{Cumulative lunar basin formation ages for basins with reliable ages. The salmon and beige regions indicate the late 
veneer and late accretion epochs respectively. Ages calculated using crater statistics (relative sequence) and two different 
chronology models to determine absolute model ages: Neukum (top) and Werner (bottom). Some of the measurements have limited statistics 
and thus the derived ages have large error bars (of up to 0.1~Ga).}
\label{fig:basinages}
\end{figure}

After the lunar crust began forming near 4.42~Ga \citep{Nem09}, the decrease in the number of planetesimals is either monotonic from 
at least 4.3~Ga onwards \citep{W14} or suffered a temporary uptick caused by a possible Late Heavy Bombardment (LHB) that could have 
coincided with Nectaris basin formation \citep{B12}. { The latter requires a lunar magma ocean solidification time much 
later than current theoretical estimates \citep{Kam15}}. Based on stratigraphic relations Wilhelms (1987) lists about 30 
pre-Nectarian basins, and 12 basins formed during the Nectarian epoch. In Fig.~\ref{fig:basinages} we plot the cumulative age of 30 
lunar basins { which are reliably dated} using the chronology function of \citet{Neu01} (top) and of \citet{W14} (bottom). The ages 
of South Pole-Aitken and Nectaris basins are 4.22 $\pm$ 0.03 and 4.05 $\pm$ 0.01~Ga respectively for the former model, and 4.44 $\pm$ 
0.04 and 4.17 $\pm$ 0.02~Ga using the latter model. Thus the late accretion epoch approximately coincides with the range from South 
Pole-Aitken to Nectaris with the Werner chronology. From our N-body simulations we find that after 4.4~Ga the planetesimal population 
still decays following a stretched exponential; between 4.4~Ga and 4.2~Ga this decay mimics a regular exponential with e-folding time 
100~Myr (see Section~4). This decay rate corresponds reasonably well with that advocated by \citet{Neu01} (140~Myr), but is faster 
than 
that suggested by \citet{W14} (200~Myr). The main difference between the two applied models is that the basins were formed in periods 
stretching either over 650 or 900 million years.\\

Basin formation derived from a dynamical model and observed record depends on the assumed projectile-basin diameter relationship and 
their underlying size-frequency distribution. We follow the detailed analysis of several crater scaling laws for both smaller-sized 
and larger craters from \citet{Min15} and \citet{Joh16}. For the Moon, we employ a mean impact speed of 17 km~s$^{-1}$ and 
simple-to-complex crater diameter $D_{\rm SC}=15$~km \citep{Joh16}, and an impact angle of 45$^\circ$.\\

In our Monte Carlo impact experiments, we use an impactor size of $D>22$~km (forming a basin $D_{\rm c,fin}>300$~km) as threshold. 
Thus, whenever an object with this diameter, or larger, strikes the Moon we assume that a basin is made and we count the number of 
such basins. Using the same methodology, we also keep track of the number of small craters we produce with diameter $D_{\rm 
c,fin}=20$~km, which we shall use as an independent calibration of the decay function. Thus, in the Monte Carlo simulations, an object 
with diameter $1<D<1.1$~km is considered to have created a 20~km crater.\\

\begin{figure}[t]
\resizebox{\hsize}{!}{\includegraphics[angle=-90]{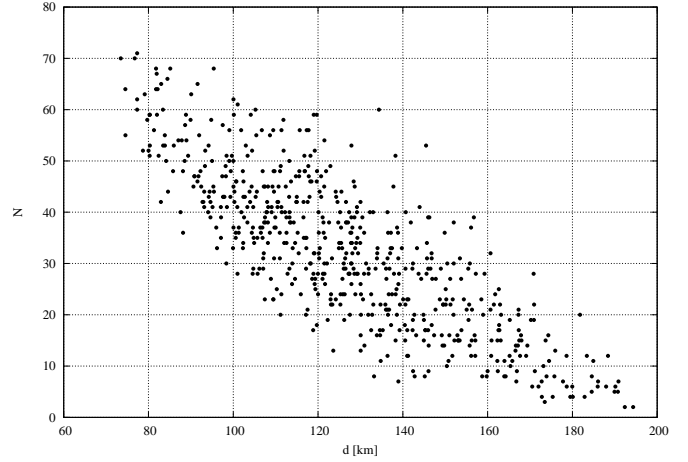}}
\caption{Number of basins produced during the late accretion epoch as a function of the size of the largest planetesimal colliding 
with the Moon. The more very large basins are produced the smaller is the total number because the total mass accreted onto the 
Moon is fixed.}
\label{fig:nbasins}
\end{figure}
In Fig.~\ref{fig:nbasins} we plot the number of basins created as a function of the diameter of the largest planetesimal that struck 
the Moon during the late accretion epoch. { Note that the Moon is not struck by any object larger than about 200~km in diameter, 
for which most collisions are accretionary \citep{R13}. Therefore our earlier assumption of `perfect' accretion is justified. From 
Fig.~\ref{fig:basinages} we can interpret the currently observed lunar basin record to be produced either between 4.3 -- 3.7~Ga with 
the Neukum chronology or between 4.4 -- 3.7~Ga employing the Werner chronology. Any basins created before 4.4~Ga are most likely 
erased by crust formation}. The number of lunar basins produced during the late accretion epoch is roughly five when using the Neukum 
chronology, so that the largest impactor should have been very large ($D>180$~km) { to be consistent with the estimated accreted 
mass}. The projectile that produced the SPA basin is thought to have had a diameter of about 170~km \citep{Pot12}. The Werner 
chronology produces ten basins during the late accretion epoch, which comports with a $D>180$~km projectile within error 
margins. { These results suggest that the South Pole-Aitken basin formed some time during or maybe just before the late accretion 
epoch, in agreement with the age estimates presented earlier.}The leftover planetesimals can provide us with 40 basins and 
the largest impactor does not exceed $\sim$140~km in diameter { during the late accretion epoch}.\\

\begin{figure}[t]
\resizebox{\hsize}{!}{\includegraphics[angle=-90]{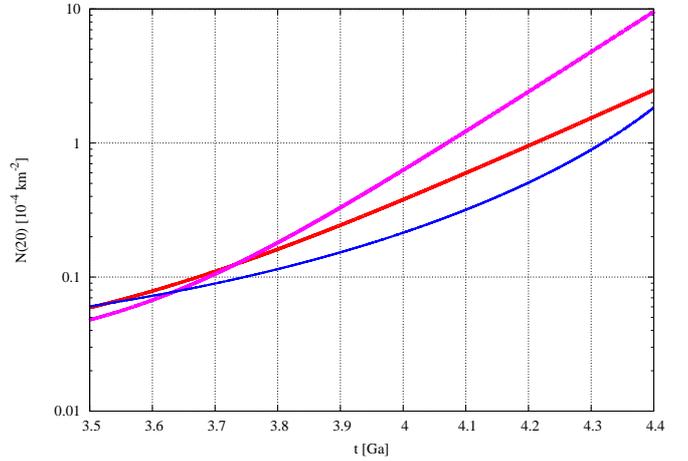}}
\caption{Chronology curves on the Moon, which show the number of 20 kilometre craters per unit area, $N(20)$ with time. We plot the 
Neukum chronology curve (magenta) \citep{Neu01}, Werner chronology curve (red) \citep{W14} and the chronology curve from the remnant 
planetesimals (blue).}
\label{fig:n20}
\end{figure}

Another test of the model is to determine the temporal cumulative density of craters with a diameter of 20~km per km$^2$. 
This quantity, $N(20)$, is calculated as follows for the Moon. Between 4.5~Ga and 4.42~Ga our Monte Carlo simulations produce on 
average 13\,000 craters with a diameter of 20~km. The crater density as a function of time becomes $N(20) = 
1.1 \times 10^{-3}$~km$^{-2}\exp[-((4.56-t)/\tau)^\beta]$, where $t$ is in Gyr, $\tau=0.012$~Gyr and $\beta=0.44$ (Section 2). The 
extrapolation to 4.56~Ga is fictitious but is needed to obtain the correct decay rate at 4.42~Ga that matches that of the N-body 
simulations. The blue curve in Fig.~\ref{fig:n20} depicts our values of $N(20)$ with age, and contains the steady-state contribution 
from the asteroid belt. The magenta curve is the Neukum chronology curve and the red curve is the Werner chronology curve, 
extrapolated from $N(1)$ where we assumed the ratio $N(1)/N(20)=1000$. Our outcome is below the Werner and Neukum chronology curves, 
indicating a deficit of craters on the Moon if the remnant planetesimals were the sole source of impactors, and suggesting the need 
for an increased bombardment rate in the more recent past { or a different size-frequency distribution at the small end 
\citep{Wer14}}.

\subsection{Hadean Earth environment}
The lower flux of planetesimals colliding with the Earth than what traditional terrestrial planet formation models predict also has 
important implications for the environment of the Hadean Earth. Our analysis shows that Earth on average accreted 0.03 wt\% between 
4.42~Ga and 4.25~Ga. This makes the Hadean Earth a much more clement environment than commonly supposed \citep{S01}. As discussed 
previously, analysis of Hadean zirons indicates that impacts were not a major source of heating and resurfacing \citep{V14,C14}, and 
oxygen isotopes in Hadean zircons suggest that liquid water existed on Earth's surface before 4.3~Ga \citep{Moj01}. The recent 
discovery of potentially biogenic carbon trapped as bona fide graphite inclusions in a 4.1~Ga zircon \citep{Bell15} yields further 
tantalising evidence that Earth could have developed a functioning biosphere before 4.1~Ga. Our results are at odds with 
\citet{Mar14}, 
who argue that between 4.15~Ga to 4.5~Ga, well over 400\% of the Hadean Earth's surface is buried by impact-generated melt, which 
would have also boiled the oceans away \citep{S89}. We would argue that in future works addressing the Hadean Earth environment, only 
a low impact flux should be considered.

\subsection{Terrestrial planet formation models}
{ Our Monte Carlo simulations indicate that there was generally} a low amount of mass in remnant planetesimals { at the time of 
the GI.} This requires that either there were never many (large) planetesimals to begin with, or most ground to dust through 
collisional cascades during the assembly of the terrestrial planets. The role and onset of collisional evolution is so far little 
explored using N-body simulations, because of inherent difficulties. The initial conditions of many of these simulations are composed 
of an unperturbed disc of planetary embryos embedded in a swarm of planetesimals (e.g. \citet{Mat16}) { and the latter would quickly 
evolve to dust \citep{KT10}}. Our result begs the question whether or not these initial conditions are realistic, because one would 
expect that the planetesimals are ground away on the same time scale that planetary embryos can form. Clearly further study is needed 
to determine whether these initial conditions are feasible and what the role of collisions are in growing terrestrial planets.\\

Pebble accretion, in which planetesimals are rapidly grown to the size of planetary embryos through the accretion of pebbles that 
drift towards the sun through the circum-solar nebula \citep{LJ12}, does not require a large planetesimal mass. It is possible to use 
pebble accretion combined with a very low ($<10^{-3}$~$M_\oplus$) total mass in planetesimals to reproduce the terrestrial planet 
system \citep{Lev15}. This setup never had a high mass in planetesimals to begin with because almost all of the accreted mass on the 
planets comes from the pebbles. Even though the results of \citet{Lev15} look like a promising alternative to the traditional 
oligarchic growth models, pebble accretion is not without its own set of issues, one of which is that there { appear to be 
insufficient} planetesimals to damp the orbits of the terrestrial planets to their current values. Collisional fragmentation from 
giant impacts may alleviate this problem \citep{G15}, but only if these planetesimals are cleared on a short time scale, and that 
their combined mass is low enough after the GI to be consistent with our above estimates. Future studies on rocky planet formation 
that rely on pebble accretion may require the inclusion of collisional fragmentation when two embryos collide. 
\\

Our results also indicate that relying on late accretion of planetesimals on the Earth coupled with dynamical simulations to 
pinpoint the timing of the Moon-forming event \citep{J14} cannot work. First, the required amount of mass accreted by the Earth 
after the GI requires a remnant mass in planetesimals that far exceeds our estimate from Section~4. Second, we have demonstrated that 
a rain of planetesimals on the Earth that reproduce the required accreted mass is inconsistent with the geochronology and HSE ratio.

\section{Conclusions}
Here, we determined amount of mass accreted by the terrestrial planets after the Moon-forming event. We compared the outcomes of 
N-body simulations of the dynamical evolution of the terrestrial planets and remnant planetesimals with Monte Carlo simulations of 
impacts on the Earth, Mars and Moon. By combining the outcome of these simulations with the lunar HSE budget, cratering 
record, geochronology and geochemistry, it is shown that

\begin{enumerate}
 \item from HSE analysis the total mass in planetesimals at the time of the Moon-forming event had to be $\sim10^{-3}$~$M_\oplus$ 
(Section 2);
{ \item the expected chondritic contribution to Mars during the late veneer epoch is 0.06 wt\% (Section 5);
 \item the high terrestrial HSE budget was most likely caused by a singular event unique to Earth (Section 6.1);
 \item the remnant planetesimals from terrestrial planet formation may account for most lunar basins, but not necessarily most of the 
small craters (Section 6.2);}
 \item the surface conditions of the Hadean Earth were much more clement than commonly thought because of a low-intensity 
bombardment (Section 6.3);
 \item new terrestrial planet formation models will need to take collisional evolution into account and explore alternatives to 
oligarchic growth such as pebble accretion (Section 6.4).
\end{enumerate}

\section{Acknowledgements}
We thank Laurie Reisberg for reminding us about osmium isotopes in martian meteorites. Norm Sleep and two anonymous 
reviewers provided us with valuable feedback that substantially improved the paper. RB is grateful for financial support from the 
Daiwa Anglo-Japanese Foundation, the Astrobiology Center Project of the National Institutes of Natural Science (AB271017), and JSPS 
KAKENHI (16K17662). RB and SJM acknowledge the John Templeton Foundation - FfAME Origins program in the support of CRiO. SJM is 
grateful for support by the NASA Exobiology Program (NNH14ZDA001N-EXO). SCW is supported by the Research Council of Norway (235058/F20 
CRATER CLOCK) and through the Centres of Excellence funding scheme, project 223272 (CEED). Numerical simulations were in part carried 
out on the PC cluster at the Center for Computational Astrophysics, National Astronomical Observatory of Japan.

Wilhelms, D. E. 1987. The Geologic History of the Moon. USGS Professional Paper 1348, 1-337. 
\end{document}